\documentclass[onecolumn,english,reprint, longbibliography, superscriptaddress, breaklinks=true, showkeys, showpacs=false, nofootinbib]{revtex4}
\usepackage[T1]{fontenc}
\usepackage[latin9]{inputenc}
\usepackage{color}
\usepackage{babel}
\usepackage{amsmath}
\usepackage{amssymb}
\usepackage{subfigure}
\usepackage{graphicx}
\usepackage{physics} 
\usepackage{quantikz}
\usepackage{ulem}
\usepackage{hyperref}
\usepackage{comment}

\usepackage{booktabs} 
\usepackage{siunitx}  

\usepackage{listings}
\usepackage{xcolor} 
\definecolor{LightGray}{gray}{0.9}
\definecolor{DarkGreen}{rgb}{0,0.5,0}

\usepackage{hyperref}
\hypersetup{
    colorlinks=true,
    linkcolor=blue,
    filecolor=blue,      
    urlcolor=blue,
    }
\makeatletter
\@ifundefined{textcolor}{}
{%
 \definecolor{BLACK}{gray}{0}
 \definecolor{WHITE}{gray}{1}
 \definecolor{RED}{rgb}{1,0,0}
 \definecolor{GREEN}{rgb}{0,1,0}
 \definecolor{BLUE}{rgb}{0,0,1}
 \definecolor{CYAN}{cmyk}{1,0,0,0}
 \definecolor{MAGENTA}{cmyk}{0,1,0,0}
 \definecolor{YELLOW}{cmyk}{0,0,1,0}
}
\pdfoutput=1
\hypersetup{colorlinks=true,citecolor=blue,linkcolor=cyan,urlcolor=blue,filecolor= green, breaklinks=true}
\usepackage{url}
\usepackage{breakurl}
\makeatother
\begin{document}

\author{Lucas Friedrich\href{https://orcid.org/0000-0002-3488-8808}{\includegraphics[scale=0.05]{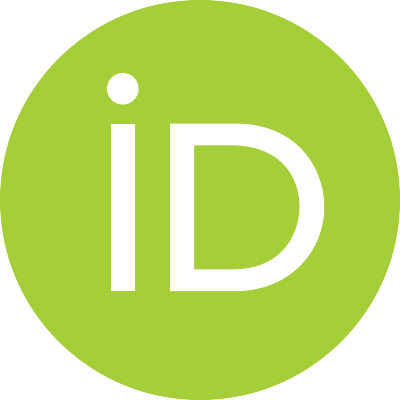}}}
\email[Electronic address: ]{lucas.friedrich@acad.ufsm.br}
\address{Physics Department, 
Federal University of Santa Maria, 97105-900,
Santa Maria, RS, Brazil}

\author{Douglas F. Pinto\href{https://orcid.org/0000-0002-3991-6935}{\includegraphics[scale=0.05]{orcidid.pdf}}}
\email{douglasfpinto@gmail.com}
\address{Physics Department, 
Federal University of Santa Maria, 97105-900,
Santa Maria, RS, Brazil}

\author{Diego S. Starke\href{https://orcid.org/0000-0002-6074-4488}{\includegraphics[scale=0.05]{orcidid.pdf}}}
\email{starkediego@gmail.com}
\affiliation{Physics Department, 
Federal University of Santa Maria, 97105-900,
Santa Maria, RS, Brazil}

\author{Jonas Maziero\href{https://orcid.org/0000-0002-2872-986X}{\includegraphics[scale=0.05]{orcidid.pdf}}}
\email[Eletronic address (corresponding author): ]{jonas.maziero@ufsm.br}
\address{Physics Department, 
Federal University of Santa Maria, 97105-900,
Santa Maria, RS, Brazil}

\title{Preparing general mixed quantum states on quantum computers}
 
\begin{abstract}
The preparation of quantum states is a fundamental subroutine for a broad class of quantum information protocols and is critical for both quantum communication and quantum computation.
Building upon the quantum algorithms introduced in previous works [M. B. Pozzobom and J. Maziero, Quantum Inf. Process. 18, 142 (2019)] and [E. R. G\aa rding \textit{et al.}, Entropy 23, 797 (2021)], the authors of [F. Shahbeigi, M. Karimi and V. Karimipour, Phys. Scr. 97, 025101 (2022)] demonstrated the capability to prepare mixed two-qubit X-real states on quantum computers by extending the methodology originally devised for mixed two-qubit Bell diagonal states. In this article, we delve into an overlooked pattern within these quantum circuits, allowing us to  
present a modular algorithm  
for the preparation of general $d$-dimensional mixed quantum states using quantum information processors. 
Our general algorithm has a modular structure, encompassing eigenvalue encoding, entropy injection, and eigenvector preparation.
To validate  our algorithm, we conduct tests on quantum computers utilizing both X- and non X-states for two mixed-state qubits, two-ququart Bell-diagonal states, as well as arbitrary random density matrices spanning one, two, and three qubits.
\end{abstract}

\keywords{Quantum simulation; Quantum computer; Mixed state preparation; Quantum resources }

\maketitle

\section{Introduction}

The first ideas about quantum computation were introduced around four decades ago by Richard Feynman, Paul Benioff, David Deutsch, and others \cite{Feynman1982,Benioff1982,Deutsch1985}. 
Since then, important developments have been made on the theoretical and hardware fronts \cite{Steane1988,Ladd2010,OBrien2007,Mohseni2024}. 
Advances in the implementation of quantum error correction codes allow us to envisage the availability of fault-tolerant quantum computers for the next decades \cite{Girvin2023, Acharya2025, Putterman2025, Reichardt2025}. In this epoch of transition from the Noisy Intermediate Scale Quantum (NISQ) era to the fault-tolerant era, we will witness an increased investigation of applications of quantum computation in diverse areas, such as optimization and artificial intelligence \cite{Biswas2017,Acampora2025}, quantum chemistry \cite{Cao2019}, molecular sciences \cite{Ollitrault2021,Liu2022,Baiardi2023}, high-energy physics \cite{Meglio2024}, and differential equations, fluid dynamics, and engineering \cite{Tennie2025}.

To enable near-term and future applications of quantum computers, it is essential to develop quantum algorithms for the more diverse types of problems \cite{Childs2010,Bharti2022,Santoro2006,Dalzell2025,Bauer2020,Montanaro2016}.
These developments, coupled with the accessibility of open-source software platforms like Qiskit \cite{qiskit}, which facilitate simulations and demonstration experiments on real IBM and other quantum devices \cite{ibmq}, have motivated the evolution of quantum algorithms for quantum state preparation. This evolution spans theoretical exploration, simulation-based investigations, and experimental applications alike \cite{Plesch2011,Cruz2019,Araujo2021,Zhang2022,Yeo2025,Yin2025,Yuan2023,Perdomo2025}.

In this context, subroutines built upon quantum state preparation (QSP) algorithms have found utility in a plethora of quantum applications \cite{kitaev1995,Shende2006,Plesch2011,Arrazola2019,Araujo2021,He2021,Zhang2021,Veras2022}. These applications span from the implementation of the general quantum Fourier transform \cite{kitaev1995}, to the simulation of noisy quantum channels \cite{Wei2018,Xin2017,Zanetti2023}, algorithms designed for the computation of expectation values in quantum systems operating at finite temperatures \cite{Clemente2024}, and the emulation of general quantum measurements \cite{Yordanov2019,Pinto2023}, all of which are instrumental in probing the properties of quantum correlations.
Moreover, the category of two-qubit Bell-diagonal states plays a pivotal role in advancing our understanding of quantum correlations and other quantum resources \cite{Caves2010}. This recognition has spurred the development of specialized preparation algorithms adapted to these states within the realm of quantum computing \cite{Pozzobom2019, Garding2021}. Subsequently, these results have been extended to encompass the broader domain of the two-qubit X-real state class \cite{Shahbeigi2022}.
This kind of density matrix has non-null elements only in the main and secondary diagonals, motivating the name X state.

Building upon the insights gleaned from prior studies, in this article we 
uncover a pattern hitherto unexplored in  these quantum circuits, thereby extending their applicability to general cases. We report a modular algorithm 
for the preparation of any mixed quantum state using quantum processors. 
Our approach splits the mixed state preparation into three layers: eigenvalues codification in
a pure state with real coefficients, entropy injection through correlations with auxiliary qubits, and the eigenvectors encoding. Besides providing explicit quantum circuits to prepare mixed two-qubit
states that have not been reported in the literature, our modular algorithmic structure can motivate the independent development of more efficient algorithms to prepare real coefficient pure states (to encode
the mixed state eigenvalues) \cite{Falco2026, Alhajjar2023, Grover2002} and to prepare certain types of eigenvectors (for particular classes of states
and for particular classes of Hamiltonians in the case of thermal states) \cite{Lomwel2026, Rouze2026}.
We evaluate the efficacy of this algorithm across a spectrum of scenarios, ranging from mixed states of two qubits, encompassing both X- and non X-states, to two-ququart Bell-diagonal states, and evaluating its performance on random density matrices spanning one, two, and three qubits. 

The subsequent sections of this article are structured as follows. In Sec.~\ref{sec:simul}, we elucidate our mixed state preparation protocol. 
In Sec.~\ref{sec:appA}, we review an algorithm we use for generating $n$-qubit pure states with real coefficients.
In Sec. \ref{sec:results} we present the quantum circuits and/or the results of our algorithm applied to two-qubit Bell-diagonal states (Sec. \ref{app_BDS}), to two-qubit X real states and non-X real states (Sec. \ref{subsec:xstates}), to two-qubit X complex states (Sec. \ref{app_XC}), to two-ququart Bell-diagonal states (Sec. \ref{app:ququart}), and to random states of one, two and three qubits (Sec. \ref{subsec: random states}).
In Sec.~\ref{sec:conc}, we 
summarize the key findings and implications of our study.
In Appendix \ref{sec:appC}, we show the calibration data of the quantum chip used in our experiments. In Appendix \ref{sec:appB}, we provide additional fidelity and Frobenius distance results for the X states and non-X states for two qubits.

\section{Mixed state preparation protocol}
\label{sec:simul}

In this section, we present in detail the operation of the mixed-state preparation algorithm. To this end, let us consider a density operator of the form:
\begin{equation}
    \rho = \sum_{j=0}^{d-1} r_j |r_j\rangle \langle r_j|,\label{eq:rho_geral}
\end{equation}
where $d = 2^n$ and $n \in \mathbb{N}$ represent the number of qubits required to prepare $\rho$. Based on this definition, the algorithm can be described through the following steps:
\begin{enumerate}
    \item Initially, we consider a quantum circuit composed of $2n$ qubits in the $|0\rangle$ state, that is:
    \begin{equation}
        |\psi_1\rangle = |0\rangle^{\otimes 2n} = |0\rangle^{\otimes n} \otimes |0\rangle^{\otimes n}.
    \end{equation}

    \item \textbf{Eigenvalue encoding.} Next, a unitary operation $U_{r_j}$ is applied to the first $n$ qubits, resulting in:
    \begin{equation}
        |\psi_2\rangle = U_{r_j}|0\rangle^{\otimes n} \otimes |0\rangle^{\otimes n} = \sum_{j=0}^{d-1} \sqrt{r_j} |j\rangle \otimes |0\rangle^{\otimes n},
    \end{equation}
    that is, the operation $U_{r_j}$ encodes the eigenvalues of $\rho$ into the state of the first $n$ qubits.

    \item \textbf{Entropy injection.} Then, a sequence of $n$ CNOT gates is applied between the first $n$ qubits (as controls) and the remaining $n$ qubits (as targets), yielding the state:
    \begin{equation}
        |\psi_3\rangle = C_{X}^{\otimes n} |\psi_2\rangle = \sum_{j=0}^{d-1} \sqrt{r_j} |j\rangle \otimes |j\rangle.
    \end{equation}
    These controlled operations are responsible for inserting entropy into the main register.
    \item \textbf{Eigenvector preparation.} Finally, a unitary operation $U_{|j\rangle \rightarrow |r_j\rangle}$ is applied to the first $n$ qubits, so that:
    \begin{equation}
        |\psi_4\rangle = (U_{|j\rangle \rightarrow |r_j\rangle} \otimes \mathbb{I}^{\otimes n}) |\psi_3\rangle = \sum_{j=0}^{d-1} \sqrt{r_j} |r_j\rangle \otimes |j\rangle,\label{eq:purification_of_rho}
    \end{equation}
    that is, the operation $U_{|j\rangle \rightarrow |r_j\rangle}$ transforms the computational basis $\{|j\rangle\}_{j=0}^{d-1}$ of the first $n$ qubits into the eigenbasis of $\rho$, namely, $\{|r_j\rangle\}_{j=0}^{d-1}$.
\end{enumerate}

Note that, after performing the steps described above, the final state of the system is given by $|\psi_4\rangle$. This state constitutes a purification of the density operator $\rho$, in the sense that tracing out the last $n$ qubits of the circuit yields exactly the desired mixed state:
\begin{equation}
\Tr_{n,\cdots,2n-1}\left(|\psi_4\rangle\langle\psi_4|\right) = \rho.
\end{equation}

\begin{figure}[t]
\centering
\includegraphics[scale=0.57]{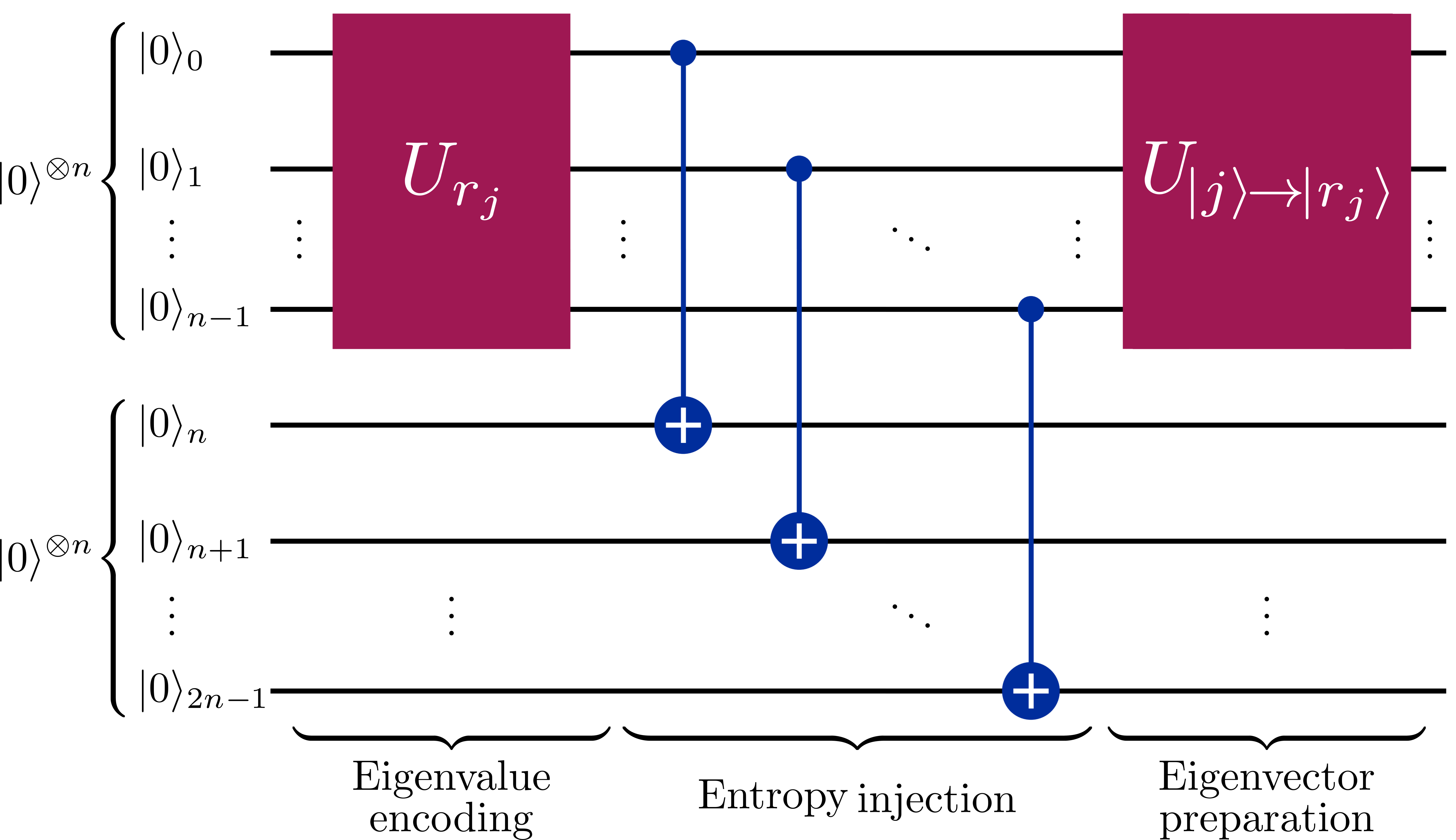}
\caption{Illustration of the quantum circuit that implements our modular algorithm used to prepare general mixed quantum states of $n$ qubits.
}
\label{fig:circuito_quantico}
\end{figure}

In Fig.~\ref{fig:circuito_quantico}, we present a schematic representation of the algorithm using quantum circuit notation. Additionally, the Qiskit code that implements the algorithm is provided in List.~\ref{lst:purification_code}, aiming to facilitate practical understanding. Also with that aim, in the next section,
we detail the implementation of each part of the algorithm when applied to some particular classes of mixed states.

\begin{lstlisting}[
    caption = {Code in Qiskit used to generate the purification of $\rho$ described by Eq.~\eqref{eq:purification_of_rho}. In this code we assume that $\rho$, defined as \texttt{rho}, was previously defined. The \texttt{initialize} function implements the algorithm for the preparation of real-coefficients pure states described in the next subsection.
    %in Appendix \ref{sec:appA}.
    },
    label = {lst:purification_code}
]
# Import the QuantumCircuit class from Qiskit to create quantum circuits
from qiskit import QuantumCircuit

# Import NumPy for numerical operations
import numpy as np

# Import eigh from SciPy to compute eigenvalues and eigenvectors of a Hermitian matrix
from scipy.linalg import eigh

# Import the Operator class from Qiskit to represent unitary operations
from qiskit.quantum_info import Operator

# Compute the eigenvalues and eigenvectors of the density matrix rho
eigenvalues, eigenvectors = eigh(rho)

# Create a quantum circuit with 2n qubits
qc = QuantumCircuit(2 * n)

# Initialize the first n qubits in a superposition state
qc.initialize(np.sqrt(eigenvalues), list(range(n)))

# Apply CNOT gates between system and ancilla qubits
for i in range(n):
    qc.cx(i, i + n)

# Create a unitary from the eigenvectors and apply it
U = Operator(eigenvectors)
qc.unitary(U, list(range(n)))
\end{lstlisting}

Finally, it is worth noting that this description is not limited to states defined over $n$ qubits. It can actually be generalized to the case in which $n$ qudits of dimension $d'$ are used. In other words, the algorithm can be applied to states $\rho$ as described in Eq. \eqref{eq:rho_geral}, but now with $d = d'^n$ instead of $d = 2^n$.

\subsection{Algorithm for preparing $n$-qubit pure states with real coefficients}
\label{sec:appA}

Here we present the algorithm utilized for the preparation of quantum states with real amplitudes, based on the methodology outlined in Ref.~\cite{Shende2006}. To elucidate the core concepts of the algorithm, we initially consider a state of a qubit:
\begin{align}
|\psi\rangle & = |c_{0}|e^{i\phi_{0}}|0\rangle + |c_{1}|e^{i\phi_{1}}|1\rangle  = e^{it/2}\big(e^{-i\phi/2}\cos(\theta/2)|0\rangle + e^{i\phi/2}\sin(\theta/2)|1\rangle\big).
\end{align}
Given that our algorithm operates with real amplitudes for the pure states, we set $\phi=0$, $t = 0$, and $\theta = \arccos(|c_0|)\in[0,\pi]$. 
Regarding the representation of this state on the Bloch sphere \cite{Nielsen2000}, it becomes evident how to align this state with the $z$ axis through the rotations $R_{z}(\phi) = \begin{bmatrix} e^{-i\phi/2} & 0 \\ 0 & e^{i\phi/2} \end{bmatrix}$ and $R_{y}(\theta) = \begin{bmatrix} \cos(\theta/2) & -\sin(\theta/2) \\ \sin(\theta/2) & \cos(\theta/2) \end{bmatrix}$. Consequently, we can express:
\begin{equation}
    |\psi\rangle = R_{y}(\theta)|0\rangle.
\end{equation}

This building block shall also be applied for more qubits. Let us now regard explicitly the two-qubit case, whose state can be written as follows:
\begin{align}
|\psi\rangle & = \sum_{j,k=0}^{1}c_{jk}|jk\rangle  = \sum_{j=0}^{1}r_{j}|j\rangle r_{j}^{-1}\sum_{k=0}^{1}c_{jk}|k\rangle \\
& = \sum_{j=0}^{1}r_{j}|j\rangle\otimes U_{j}|0\rangle   = C_{U_0}^{0_0\rightarrow 1}C_{U_1}^{0_1\rightarrow 1}\big(R_y(\xi)|0\rangle\otimes|0\rangle\big)
\end{align}
where $U_j = R_{y}(\theta_j)$ with $\theta_j = 2\arctan(|c_{j1}|/|c_{j0}|)$, $\xi=2\arccos(r_0)$, and $r_j^2 = |c_{j0}|^{2} + |c_{j1}|^{2}$. Above we used the controlled unitary $C_{U}^{c_s \rightarrow t}$ with $c$ standing for the control qubit, $t$ is the target qubit and $s$ is the activation state. For instance $C_{U}^{1_0 \rightarrow 0} = U\otimes|0\rangle\langle 0| + \mathbb{I}\otimes|1\rangle\langle 1|.$ Besides, we notice that $R_y (\xi)|0\rangle = \sum_{j=0}^{1}r_j |j\rangle$ is a general one-qubit state with real coefficients, when represented in the computational basis (CB). This pattern shall repeat for an $n$-qubit state, that is prepared starting from a $n-1$ state with real coefficients in the CB.

For $3$-qubit states
\begin{align}
|\psi\rangle & = \sum_{j,k,l=0}^{1}c_{jkl}|jkl\rangle = \sum_{j,k=0}^{1}r_{jk}|jk\rangle r_{jk}^{-1}\sum_{l=0}^{1}c_{jkl}|l\rangle \\
& = \sum_{j,k=0}^{1}r_{jk}|jk\rangle\otimes U_{jk}|0\rangle = \Pi_{j,k=0}^{1} C_{U_{jk}}^{0_{j}1_{k}\rightarrow 2}\big(|\Phi\rangle\otimes|0\rangle\big)
\end{align}
with $U_{jk} = R_{y}(\theta_{jk})$,
$\theta_{jk} = 2\arctan\big(|c_{jk1}|/|c_{jk0}|\big)$, $r_{jk}^2 = |c_{jk0}|^2 + |c_{jk1}|^2$ and $|\Phi\rangle = \sum_{j,k=0}^{1}r_{jk}|jk\rangle$ is a two-qubit state with real coefficients.

So, in the general case of $n$ qubits, the state preparation will proceed by the following steps:
\begin{enumerate}
\item Prepare a $(n-1)$-qubit state
\begin{equation}
|\Phi\rangle = \sum_{j_0,j_1,\cdots,j_{n-2}=0}^{1}r_{j_0,j_1,\cdots,j_{n-2}}|j_0,j_1,\cdots,j_{n-2}\rangle
\end{equation}
with real coefficients 
\begin{equation}
r_{j_0,j_1,\cdots,j_{n-2}} = \sqrt{|c_{j_0,j_1,\cdots,j_{n-2},0}|^{2} + |c_{j_0,j_1,\cdots,j_{n-2},1}|^{2}}.
\end{equation}

\item Apply $2^{n-1}$ multi-controlled unitary gates to $|\Phi\rangle\otimes|0\rangle$ with the first $n-1$ qubits as the control register and the last qubit as the target:
\begin{equation}
\Pi_{j_0,j_1,\cdots,j_{n-2}=0}^{1}C_{U_{j_0,j_1,\cdots,j_{n-2}}}^{0_{j_0},1_{j_1},\cdots,(n-2)_{j_{n-2}}\rightarrow n-1}
\end{equation}
with
\begin{align}
& U_{j_0,j_1,\cdots,j_{n-2}} = R_{y}(\theta_{j_0,j_1,\cdots,j_{n-2}}),
\end{align}
where
\begin{align}
    \theta_{j_0,j_1,\cdots,j_{n-2}} = 2\arctan\left(\frac{|c_{j_0,j_1,\cdots,j_{n-2},1}|}{|c_{j_0,j_1,\cdots,j_{n-2},0}|}\right),
\end{align}
\end{enumerate}
which conclude the method for eigenvalue encoding.

\section{Results}
\label{sec:results}

In this section, we present several application examples of the mixed-state preparation algorithm proposed in this work. In Sec. \ref{app_BDS}, we instantiate the main steps of our protocol using two-qubit Bell-diagonal states. In Sec. \ref{subsec:xstates}, we analyze the performance of the algorithm when preparing X- and non X-states of two qubits.  
In Sec. \ref{app_XC}, we detail the quantum circuits for the preparation of Bell-diagonal and complex X-states.
In Sec. \ref{app:ququart}, we give the quantum circuit for the preparation of two-ququart Bell-diagonal states.
In Sec. \ref{subsec: random states}, we explore our algorithm applied in the preparation of general states involving one, two, and three qubits.
All experiments were performed on the IBM Q Kingston device, whose calibration data are provided in Appendix~\ref{sec:appC}.

\subsection{Algorithm for preparation of Bell-diagonal states}
\label{app_BDS}

Bell diagonal states (BDS) are two-qubit density matrices that have the four Bell states as eigenvectors: 
\begin{equation}
\rho_{bd} = \sum_{j_0,j_1=0}^1 p_{j_0 j_1}|\Phi_{j_0 j_1}\rangle\langle \Phi_{j_0 j_1}|,
\end{equation}
with $\{p_{j_0 j_1}\}$ being a probability distribution and the Bell states are given in terms of the computational basis as $|\Phi_{j_0 j_1}\rangle = 2^{-1/2}\sum_{l=0}^1 e^{2\pi ij_1 l/2}|(l+j_0)\bmod 2\rangle\otimes|l\rangle$.

The steps of our algorithm applied in this particular case are as follows: \\

\textbf{1st step}: Prepare all four qubits in the standard state $|0\rangle$.

\vspace{0.2cm}

\textbf{2nd step}: Apply $U_{r_j}$ to ``prepare'' the eigenvalues of $\rho$. Two-qubit states have four eigenvalues, that can be encoded in a pure two-qubit state using an $R_y$ gate followed by two controlled $R_y$ gates:
\begin{align}
|\Psi_{r_j}\rangle & = C_{R_y(\theta_0)}^{0_0\rightarrow 1}C_{R_y(\theta_1)}^{0_1\rightarrow 1}\big(R_y(\xi)|0\rangle\otimes|0\rangle\big) \nonumber\\
 & = \cos(\xi/2)\cos(\theta_0/2)|00\rangle + \cos(\xi/2)\sin(\theta_0/2)|01\rangle  + \sin(\xi/2)\cos(\theta_1/2)|10\rangle + \sin(\xi/2)\sin(\theta_1/2)|11\rangle \nonumber\\
 & = \sum_{j_0,j_1=0}^1 \sqrt{p_{j_0 j_1}}|j_0 j_1\rangle.
\end{align}

\textbf{3rd step}: Add two ancilla qubits in the state $|0\rangle$ and apply controlled NOT gates with the ancilla as target qubits:
\begin{align}
|\Psi_p\rangle & = C_X^{0_1\rightarrow 2}C_X^{1_1\rightarrow 3}|\Psi_{r_j}\rangle\otimes|00\rangle \nonumber \\
& = \sum_{j_0,j_1=0}^1 \sqrt{p_{j_0 j_1}}|j_0 j_1\rangle\otimes|j_0j_1\rangle.
\end{align}
This operation injects entropy into the first two qubits due to the correlations created with the auxiliary qubits.

\textbf{4th step}: Change from the computational basis to the eigenbasis of $\rho$, which in this case is the Bell basis. In this case, it is well known that the basis change is achieved by a Hadamard gate on the first qubit followed by a controlled NOT gate, i.e.,
\begin{align}
  |\Psi_{|r_j\rangle}\rangle & = U_{|j\rangle\rightarrow|r_j\rangle}|\Psi_p\rangle \nonumber \\
  & = C_{X}^{0_1\rightarrow 1}(H\otimes \mathbb{I})\sum_{j_0,j_1=0}^1 \sqrt{p_{j_0 j_1}}|j_0 j_1\rangle\otimes|j_0j_1\rangle \nonumber \\
  & = \sum_{j_0,j_1=0}^1 \sqrt{p_{j_0 j_1}}|\Phi_{j_0j_1}\rangle\otimes|j_0 j_1\rangle.
\end{align}
With this, tracing out the ancilla qubits, we get
\begin{equation}
\Tr_{23}|\Psi_{|r_j\rangle}\rangle\langle\Psi_{|r_j\rangle}| = \sum_{j_0,j_1=0}^1 p_{j_0 j_1}|\Phi_{j_0j_1}\rangle\langle\Phi_{j_0j_1}| \equiv \rho_{bd}.
\end{equation}
The corresponding quantum circuit is shown in Fig.~\ref{qc_bds}.

\begin{widetext}
\begin{center}
\begin{figure}[ht]
\centering
\includegraphics[scale=0.75]{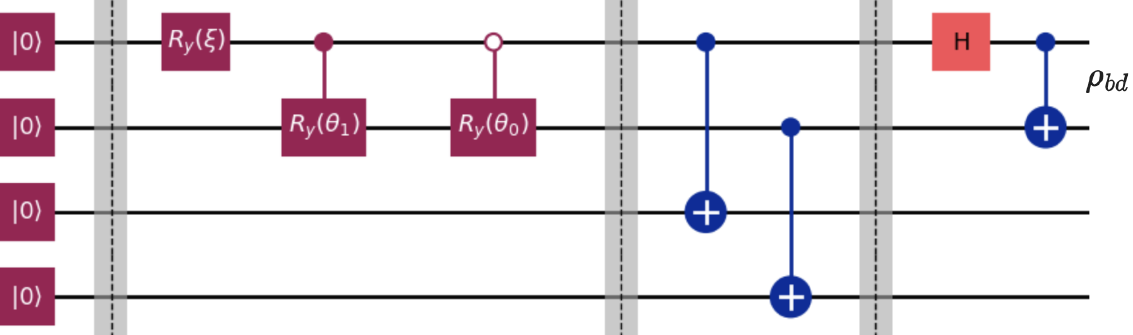}
\caption{Quantum circuit for preparing two-qubit Bell diagonal states. The second block ``prepares'' the eigenvalues of a mixed two-qubit state. The third block injects entropy through entanglement with the auxiliary qubits. The fourth block changes from the computational to the basis of eigenvectors of the density operator.}
\label{qc_bds}
\end{figure}
\end{center}
\end{widetext}

The quantum circuit above would be a useful tool for verifying, for example, the entanglement properties of the Gibbs thermal state associated with magnetic dipolar interaction Hamiltonian \cite{Castro2016}.

\subsection{Preparing X-states and non X-states}
\label{subsec:xstates}

\begin{figure}[t]
\centering
\includegraphics[width=0.6\linewidth]{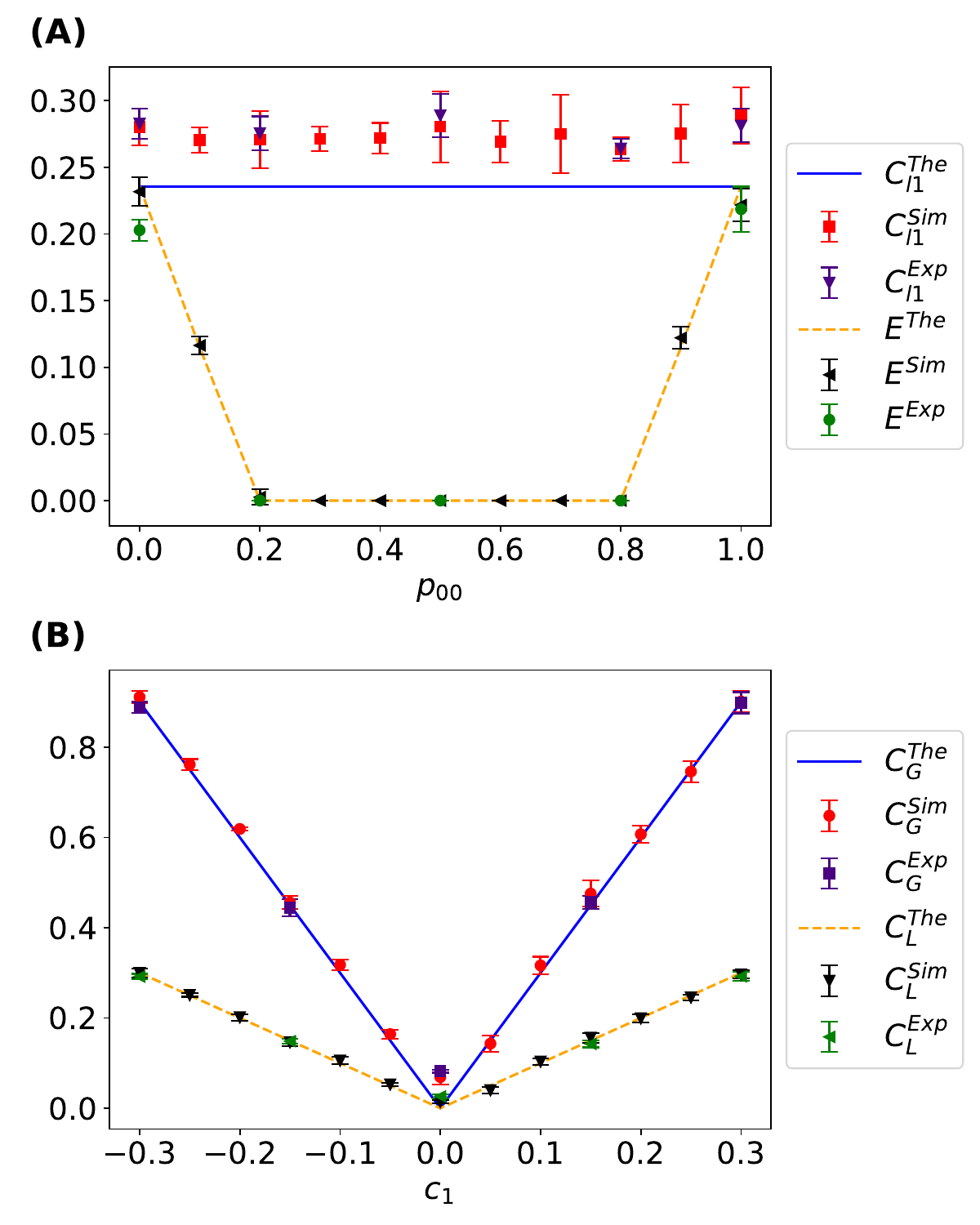}
\caption{
\textbf{(A)} Coherence and entanglement for the X state defined in Eq. \eqref{eq:x_state}. As a reference, we present the theoretical curves corresponding to the coherence ($C_{l1}^{\text{The}}$) and the entanglement ($E^{\text{The}}$). For each point shown in the plot, whether it refers to simulated coherence ($C_{l1}^{\text{Sim}}$), experimental coherence ($C_{l1}^{\text{Exp}}$), simulated entanglement ($E^{\text{Sim}}$), or experimental entanglement ($E^{\text{Exp}}$), the same state was prepared five times. Therefore, each point represents the average value and the standard deviation of the coherence or entanglement, allowing us to analyze, for instance, the effects of experimental noise and statistical fluctuations on the prepared state. \textbf{(B)} Plot of global coherence, indicated by the subscript G, and local coherence, indicated by the subscript L, for the non-X state defined in Eq. \eqref{eq:non_x_state}. Again, for each data point, the same state was prepared five times. Each point in the plot represents the average value and the corresponding standard deviation. For the experiments, we used the IBM quantum chip: \textit{ibm\_kingston}.
}
\label{fig:plot_coherence_global_local_and_entanglement}
\end{figure}

We begin the validation of the algorithm by considering two-qubit states. In the first example, we employ the algorithm to prepare mixed states that yield X-states in a two-qubit system \cite{Shahbeigi2022}. The obtained results are presented in Fig.~\ref{fig:plot_coherence_global_local_and_entanglement}~\textbf{(A)}. The X-real states are described by density matrices of the form:
\begin{equation}
    \rho_X  = \begin{bmatrix} 
        a & 0 & 0 & w \\ 
        0 & b & z & 0 \\ 
        0 & z & c & 0 \\ 
        w & 0 & 0 & d 
    \end{bmatrix},
\end{equation}
where $d = 1 - a - b - c$, and the elements $w$ and $z$ are assumed to be real, defining the real-valued structure of X-states. This density matrix can be diagonalized, leading to the expression:
\begin{align}
    \rho_X = \sum_{j,k=0}^{1}p_{jk}|\Psi_{jk}\rangle\langle\Psi_{jk}|,\label{eq:x_state}
\end{align}
with eigenvectors given by:
\begin{align}
    & |\Psi_{00}\rangle = \cos\theta|00\rangle+\sin\theta|11\rangle, \\
    & |\Psi_{01}\rangle = \sin\phi|01\rangle+\cos\phi|10\rangle, \\
    & |\Psi_{10}\rangle = \cos\phi|01\rangle-\sin\phi|10\rangle, \\
    & |\Psi_{11}\rangle = -\sin\theta|00\rangle+\cos\theta|11\rangle.
\end{align}

For specific angle values, such as $\theta = \phi = k\pi/4$ with $k = 1, 3, 5, 7$, the state $\rho_X$ belongs to the class of Bell-diagonal states, implying that the eigenvector basis above coincides with the Bell basis.

Next, since the algorithm is capable of generating not only X-states but also a wide variety of more general states, we present a second example in Fig.~\ref{fig:plot_coherence_global_local_and_entanglement}~\textbf{(B)}, demonstrating its application in the preparation of such more general two-qubit states. To that end, we note that the decomposition of arbitrary two-qubit density operators in the Pauli basis can be written as:
\begin{equation}
    4\rho = \sigma_{0}\otimes\sigma_{0} + \sigma_{0}\otimes\sum_{k=1}^{3}b_{k}\sigma_{k} + \sum_{j=1}^{3}a_{j}\sigma_{j}\otimes\sigma_{0} \nonumber  + \sum_{k=1}^{3}c_{k}\sigma_{k}\otimes\sigma_{k},
\end{equation}
where $\sigma_0 = |0\rangle\langle 0| + |1\rangle\langle 1|$, $\sigma_1 = |0\rangle\langle 1| + |1\rangle\langle 0|$, $\sigma_2 = -i|0\rangle\langle 1| + i|1\rangle\langle 0|$, and $\sigma_3 = |0\rangle\langle 0| - |1\rangle\langle 1|$. Based on this decomposition, we select a particular class of states defined by the conditions $c_k = a_1 = b_1 = c_1 \quad \forall k$ and $a_{2} = a_{3} = b_{2} = b_{3} = 0$. Accordingly, the density matrix takes the form:
\begin{equation}
    \rho_{c_1}= \frac{1}{4} \begin{pmatrix}
        1+c_{1} & c_{1} & c_{1} & 0 \\
        c_{1} & 1-c_{1} & 2c_{1} & c_{1} \\
        c_{1} & 2c_{1} & 1-c_{1} & c_{1} \\
        0 & c_{1} & c_{1} & 1+c_{1} 
    \end{pmatrix}.\label{eq:non_x_state}
\end{equation}

To indirectly assess the effectiveness of the proposed protocol, we first employ the $l_1$-norm coherence measure, defined as:
\begin{equation}
    C_{l_1}(\rho) = \sum_{j\ne k}|\rho_{j,k}|,
\end{equation}
where $\rho_{j,k}$ denotes the matrix elements of the density operator $\rho$ in the computational basis $\{|00\rangle, |01\rangle, |10\rangle, |11\rangle\}$.

In addition, we analyze the entanglement concurrence
\cite{Wootters2001}:
\begin{equation}
    E_{C}(\rho)=\max\left\{ 0,\sqrt{\lambda^{I}}-\sqrt{\lambda^{II}}-\sqrt{\lambda^{III}}-\sqrt{\lambda^{IV}}\right\},
\end{equation}
where $\lambda^{I}\geq\lambda^{II}\geq\lambda^{III}\geq\lambda^{IV}$ are the eigenvalues of the matrix $R = \rho\tilde{\rho}$, with $\tilde{\rho} := \sigma_{y}\otimes\sigma_{y}\rho^{*}\sigma_{y}\otimes\sigma_{y}$.

In the case of the state defined in Eq.~\eqref{eq:x_state}, we examine a mixed configuration with $\theta = \phi = \pi/8$, allowing $p_{00}$ to vary under the constraint that the eigenvalues form a valid probability distribution $\left\{\frac{p_{00}}{3}, \frac{1-p_{00}}{3}, \frac{2p_{00}}{3}, \frac{2(1-p_{00})}{3}\right\}$. This choice enables the generation of a non-Bell-diagonal X-state.

On the other hand, for the state defined by Eq.~\eqref{eq:non_x_state}, whose entanglement is zero, our analysis considers both the global and local contributions to $l_1$-norm quantum coherence. We emphasize that this class of two-qubit states was not addressed in previous studies such as Refs.~\cite{Pozzobom2019,Garding2021,Shahbeigi2022}, which were restricted to the real X-states.

Additionally, beyond coherence and entanglement, we also analyze the fidelity and the Frobenius distance with respect to the states defined in Eq. \eqref{eq:x_state} and Eq. \eqref{eq:non_x_state}. The corresponding graphs for these quantities are presented in Figs. \ref{fig:fidelity_X_state_non_X_state} and Fig.~\ref{fig:frobenius_distance_X_state_non_X_state} of Appendix~\ref{sec:appB}, respectively.

\subsection{Algorithm for preparation complex X-states}
\label{app_XC}

\begin{figure}
\centering   \includegraphics[width=0.5\linewidth]{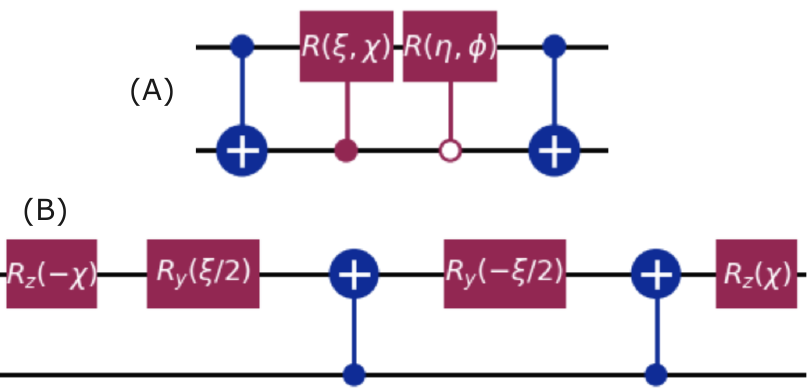}
\caption{Preparation of X complex class of two-qubit states. In (A) is shown the quantum circuit to implement the eigenvector basis related unitary $U_{|j\rangle\rightarrow|r_j\rangle}$. In (B) is shown the quantum circuit to implement the controlled rotation $C_{R(\xi,\chi)}^{1_1\rightarrow 0}$. The rest of the quantum circuit is as shown in Fig.~\ref{qc_bds}.}
\label{fig_Xc}
\end{figure}

It is worthwhile mentioning that it is only the last part of the quantum circuit in Fig.~\ref{qc_bds} that needs to be changed to prepare more general two-qubit states. This block implements the basis change using the unitary transformation 
\begin{equation}
U_{|j\rangle\rightarrow|r_j\rangle} = \begin{bmatrix} |r_0\rangle & |r_1\rangle & |r_2\rangle & |r_3\rangle\end{bmatrix}.
\end{equation}
For example, for complex X states $\rho_X = \sum_{j,k}p_{jk}|\Psi_{jk}\rangle\langle\Psi_{jk}|$ with eigenvectors $|\Psi_{00}\rangle = \cos(\eta/2)|00\rangle + e^{i\phi}\sin(\eta/2)|11\rangle,\ 
|\Psi_{01}\rangle = \cos(\xi/2)|01\rangle + e^{i\chi}\sin(\xi/2)|10\rangle, \
 |\Psi_{10}\rangle = -e^{-i\chi}\sin(\xi/2)|01\rangle + \cos(\xi/2)|10\rangle, \
 |\Psi_{11}\rangle = -e^{-i\phi}\sin(\eta/2)|00\rangle + \cos(\eta/2)|11\rangle$, we have
\begin{align}
U_{|j\rangle\rightarrow|r_j\rangle} & = \begin{bmatrix} c_\eta & 0 & 0 & -e^{-i\phi}s_\eta \\ 0 & c_\xi & -e^{-i\chi}s_\xi&0 \\ 0 & e^{i\chi}s_\xi & c_\xi &0 \\ e^{i\phi}s_\eta & 0 & 0 & c_\eta \end{bmatrix}  = C_X^{0_1\rightarrow 1}C_{R(\eta,\phi)}^{1_0\rightarrow 0}C_{R(\xi,\chi)}^{1_1\rightarrow 0}C_X^{0_1\rightarrow 1},
\end{align}
where we denoted $c_\theta := \cos(\theta/2),\ s_\theta:=\sin(\theta/2)$, with $\theta=\eta, \xi$. The controlled rotation gates above are decomposed as follows: $C_{R(\xi,\chi)}^{1_1\rightarrow 0} = \big(R_z(\chi)\otimes \mathbb{I}\big)C_X^{1_1\rightarrow 0}\big(R_y(-\xi/2)\otimes \mathbb{I}\big)C_X^{1_1\rightarrow 0}\big(R_y(\xi/2)\otimes \mathbb{I}\big)\big(R_z(-\chi)\otimes \mathbb{I}\big)$ and $C_{R(\xi,\chi)}^{1_0\rightarrow 0} = \big(\mathbb{I}\otimes X\big)C_{R(\xi,\chi)}^{1_1\rightarrow 0}\big(\mathbb{I}\otimes X\big)$, where $R(\xi,\chi)=R_z(\chi)R_y(\xi)R_z(-\chi)$. The quantum circuit to prepare the states of the X complex class is shown in Fig.~\ref{fig_Xc}.
The preparation of this more general type of X state,
\begin{equation}
\rho_X = \begin{bmatrix}
\rho^X_{00} & 0 & 0 & \rho^X_{03} \\
0 & \rho^X_{11} & \rho^X_{12} & 0 \\
0 & \rho^X_{21} & \rho^X_{22} & 0 \\
\rho^X_{30} & 0 & 0 & \rho^X_{33} \\
    \end{bmatrix}
\end{equation}
with non-null matrix elements $\rho^X_{00}=p_{00}c_\eta^2+p_{11}s_\eta^2$, $\rho^X_{03}=(\rho^X_{03})^*=(p_{00}-p_{11})e^{-i\phi}s_\eta c_\eta$, $\rho^X_{11}=p_{01}c_\xi^2+p_{10}s_\xi^2$, $\rho^X_{12}=(\rho^X_{12})^*= (p_{01}-p_{10})e^{-i\chi}s_\xi c_\xi$, $\rho^X_{22}=p_{01}s_\xi^2+p_{10}c_\xi^2$ and $\rho^X_{33}=p_{00}s_\eta^2+p_{11}c_\eta^2$, 
can be a useful tool for studying the quantum resources of thermal states generated, e.g. by the Heisenberg Hamiltonian \cite{Zhang2005}.

A minor modification of the quantum circuit in Fig.~\ref{fig_Xc} can be used to prepare a two-qubit state beyond the X form. By applying the Hadamard gate to the first qubit, it is possible to generate a more general state with all elements of the density matrix being non-zero. The unitary transformation is converted to
\begin{align}
U_{|j\rangle\rightarrow|r'_j\rangle}^{H} &=
\frac{1}{\sqrt{2}}
\begin{bmatrix}
c_\eta & - e^{-i\phi} s_\eta & c_\eta & e^{-i\phi} s_\eta \\
- e^{-i\chi} s_\xi & c_\xi & e^{-i\chi} s_\xi & c_\xi \\
c_\xi & e^{i\chi} s_\xi & -c_\xi & e^{i\chi} s_\xi \\
e^{i\phi} s_\eta & c_\eta & e^{i\phi} s_\eta & -c_\eta
\end{bmatrix}  = C_X^{0_1\rightarrow 1}C_{R(\eta,\phi)}^{1_0\rightarrow 0}C_{R(\xi,\chi)}^{1_1\rightarrow 0}C_X^{0_1\rightarrow 1} (H\otimes \mathbb{I})
\end{align}
and the density matrix takes the form
\begin{equation}
\rho = \frac{1}{2}\begin{bmatrix}
A_\eta & B & C & D\\
B^{\ast} & E_\xi & F & G\\
C^{\ast} & F^{\ast} & A_\xi & H\\
D^{\ast} & G^{\ast} & H^{\ast} & E_\eta
\end{bmatrix},
\end{equation}
with the following definitions: $A_\theta = 2(p_{00}+p_{10})c_\theta^{2} + 2(p_{01}+p_{11})s_\theta^{2},\ 
B = (p_{10}-p_{00})e^{i\chi}s_\xi c_\eta + (p_{11}-p_{01})e^{-i\phi}s_\eta c_\xi,\ 
C = (p_{00}-p_{10})c_\eta c_\xi + (p_{11}-p_{01})e^{-i(\chi+\phi)}s_\eta s_\xi,\ 
D = e^{-i\phi}s_\eta c_\eta\big(p_{00}-p_{01}+p_{10}-p_{11}\big),\ 
E_\theta = 2(p_{00}+p_{10})s_\theta^{2} + 2(p_{01}+p_{11})c_\theta^{2},\ 
F = -\,e^{-i\chi}s_\xi c_\xi \big(p_{00}-p_{01}+p_{10}-p_{11}\big),\ 
G = (p_{10}-p_{00})\,e^{-i(\chi+\phi)}s_\eta s_\xi + (p_{01}-p_{11})c_\eta c_\xi$, and $
H= (p_{00}-p_{10})e^{-i\phi}s_\eta c_\xi + (p_{01}-p_{11})e^{i\chi}s_\xi c_\eta$, with $\theta = \eta, \xi$.

\subsection{
Quantum circuit for preparing two-ququart Bell-diagonal states
}
\label{app:ququart}

For two $d$-dimensional quantum systems (qudits) $A$ and $B$, the generalization of Bell basis states is given by
\begin{equation}
    |\Phi_{jk}\rangle_{AB} = CNOT^{B\rightarrow A}_{d}\big(\mathbb{I}_A\otimes F_B\big)\big(|j\rangle_{A}\otimes|k\rangle_{B}\big),
\end{equation}
where 
\begin{equation}
F = \frac{1}{\sqrt{d}}\sum_{j,k=0}^{d-1}e^{2\pi ijk/d}|j\rangle\langle k|
\label{eq:fourier}
\end{equation}
is the quantum Fourier transform and 
$CNOT_{d}^{B\rightarrow A} = \sum_{j=0}^{d-1} X(j)\otimes|j\rangle\langle j|$
is the generalization of the controlled NOT gate for qudits. Above $X(j)=\sum_{k=0}^{d-1}|(j+ k)\bmod d\rangle\langle k|$ is state shift operator.

Two-qudit Bell-diagonal density matrices have the spectral decomposition
\begin{equation}
    \rho_{BD}^d = \sum_{j,k=0}^{d-1}p_{jk}|\Phi_{jk}\rangle_{AB}\langle\Phi_{jk}|,
\end{equation}
with $p_{jk}$ being a probability distribution.
Let us consider the preparation of two-ququart ($d=4$) Bell-diagonal density matrices. We shall encode each ququart using two qubits. The computational basis correspondence is the standard one: $\{|0\rangle,|1\rangle,|2\rangle,|3\rangle\}\rightarrow\{|00\rangle,|01\rangle,|10\rangle,|11\rangle\}$. The eigenvalues $p_{jk}$ encoding is done as explained in Appendix \ref{sec:appA}. The quantum circuit for the quantum Fourier transform in Eq. (\ref{eq:fourier}) is well known in Quantum Computation \cite{Nielsen2000}.

The last part of the algorithm is the eigenvectors encoding. This is accomplished by the unitary operation
\begin{equation}
    U_{|jk\rangle\rightarrow|\Phi_{jk}\rangle} = CNOT^{B\rightarrow A}_{4}\big(\mathbb{I}_A\otimes F_B\big).
\end{equation}
We still need to show how to apply the CNOT operation in this case. For what comes next, we use the qubits indexed $0$ and $1$ to encode the ququart $A$ and the qubits $3$ and $4$ to encode the ququart $B$.  First let us notice that this controlled operation can be decomposed as follows
\begin{equation}
    CNOT^{B\rightarrow A}_{4} = C_{X(0)}^{B_0\rightarrow A}C_{X(1)}^{B_1\rightarrow A}C_{X(2)}^{B_2\rightarrow A}
    C_{X(3)}^{B_3\rightarrow A} \equiv C_{X(1)}^{3_0 2_1\rightarrow 10}C_{X(2)}^{3_1 2_0\rightarrow 10}
    C_{X(3)}^{3_1 2_1\rightarrow 10},
    \label{eq:qc_cnot_ququart}
\end{equation}
in which we are considering qubits 1 and 3 of each of the subsystems as representing the most significant bit (MSB) and qubits 0 and 2 of each of the subsystems as the least significant bit (LSB).
Above, as $X(0)$ is the identity operator, we ignored $C_{X(0)}^{B_0\rightarrow A}$.

Observing that the state shift operators act in the following way
\begin{align}
    & X(1)|00\rangle=|01\rangle,\ X(1)|01\rangle=|10\rangle,\ X(1)|10\rangle=|11\rangle,\ X(1)|11\rangle=|00\rangle, \\
    & X(2)|00\rangle=|10\rangle,\ X(2)|01\rangle=|11\rangle,\ X(2)|10\rangle=|00\rangle,\ X(2)|11\rangle=|01\rangle, \\
    & X(3)|00\rangle=|11\rangle,\ X(3)|01\rangle=|00\rangle,\ X(3)|10\rangle=|01\rangle,\ X(3)|11\rangle=|10\rangle,
\end{align}
we notice that these operators can be implemented through the following sequence of one- and two-qubit gates: $X(1)=(\mathbb{I}\otimes X)C_X^{B\rightarrow A}$, $X(2)=X\otimes \mathbb{I}$, and $X(3)=(X\otimes X)C_X^{B\rightarrow A}$.

\begin{center}
\begin{figure}
    \centering
    \includegraphics[width=0.6\linewidth]{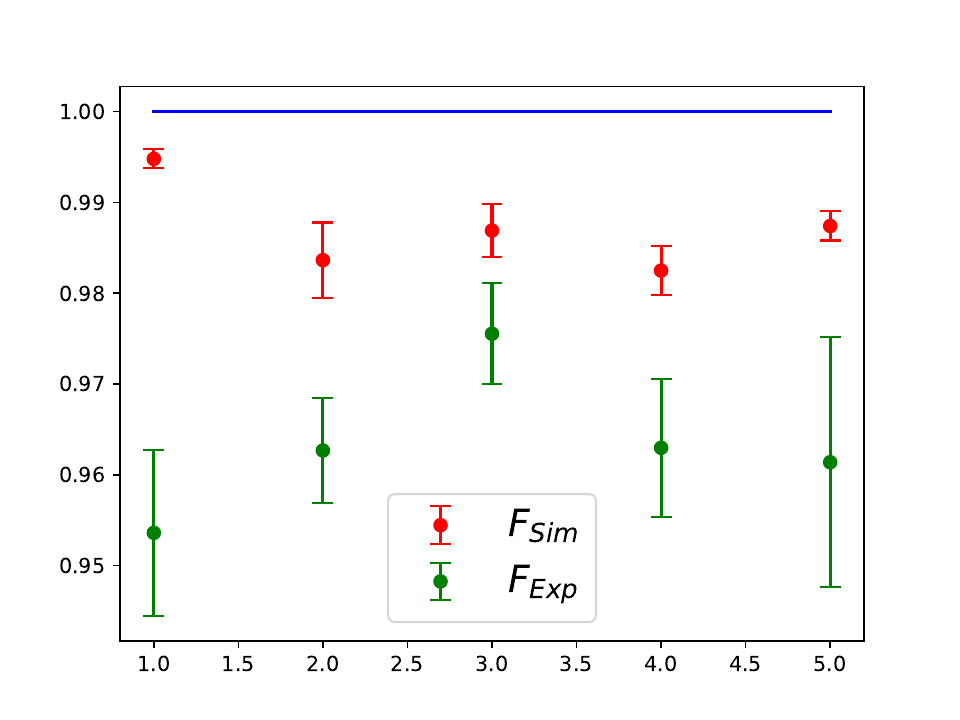}
    \caption{Analysis of simulated ($F_{Sim}$) and experimental ($F_{Exp}$) fidelity for random Bell-diagonal states of two-ququarts. For each generated state, both simulated and experimental, fidelity was evaluated five times; thus, each point in the graphs represents the average fidelity and its corresponding standard deviation.}
    \label{fig_fid_ququarts}
\end{figure}
\end{center}

On the other hand, the action of the ququart controlled-NOT operation is
\begin{align}
    & CNOT_4^{B\rightarrow A}|j\rangle|k\rangle = |(j+k)\bmod 4\rangle|k\rangle,
\end{align}
therefore,
\begin{align}
    & CNOT_4^{B\rightarrow A}|j_0 j_1\rangle_A|k_0 k_1\rangle_B = |j_0\oplus k_0\rangle |j_1\oplus k_1\oplus j_0 k_0\rangle|k_0 k_1\rangle,
\end{align}
where $j_0$ and $k_0$ are the LSB, and $j_1$ and $k_1$ are the MSB.
Above $\oplus$ stands for the sum modulo two. The final state of qubit $2$, $|j_0\oplus k_0\rangle$ is obtained using a CNOT from qubit $2$ to $0$ or vice-versa. Looking at the final state of qubit $0$, $|j_1\oplus k_1\oplus j_0 k_0\rangle$, we see that $j_1\oplus k_1$ is obtained with a CNOT from qubit $3$ to qubit $1$ or vice-versa. The carry $j_0 k_0$ (which is obtained by the ordinary operation AND) in the binary sum is then implemented using a Toffoli gate $C_X^{j_0 k_0\rightarrow j_1}$ with qubits $0$ and $2$ as controls and qubit $2$ as the target. So, related to our encoding to qubits, the operation is obtained as: 
\begin{equation}
CNOT_4^{B\rightarrow A} =C_X^{3_1\rightarrow 1}C_X^{2_1\rightarrow 0}C_X^{3_1 2_1\rightarrow 1}.
\end{equation}
With this, we obtain a shorter quantum circuit compared to Eq. (\ref{eq:qc_cnot_ququart}).
So, the quantum circuit to prepare Bell-diagonal states of two ququarts is illustrated in Fig. \ref{qc_bds_ququart}. The fidelity results for attesting the correct functioning of this quantum circuit are shown in Fig. \ref{fig_fid_ququarts}.
With this additional example, we illustrate the versatility of our modular mixed state preparation algorithm.

\begin{widetext}
\begin{center}
\begin{figure}[h]
    \centering
    \includegraphics[width=0.8\linewidth]{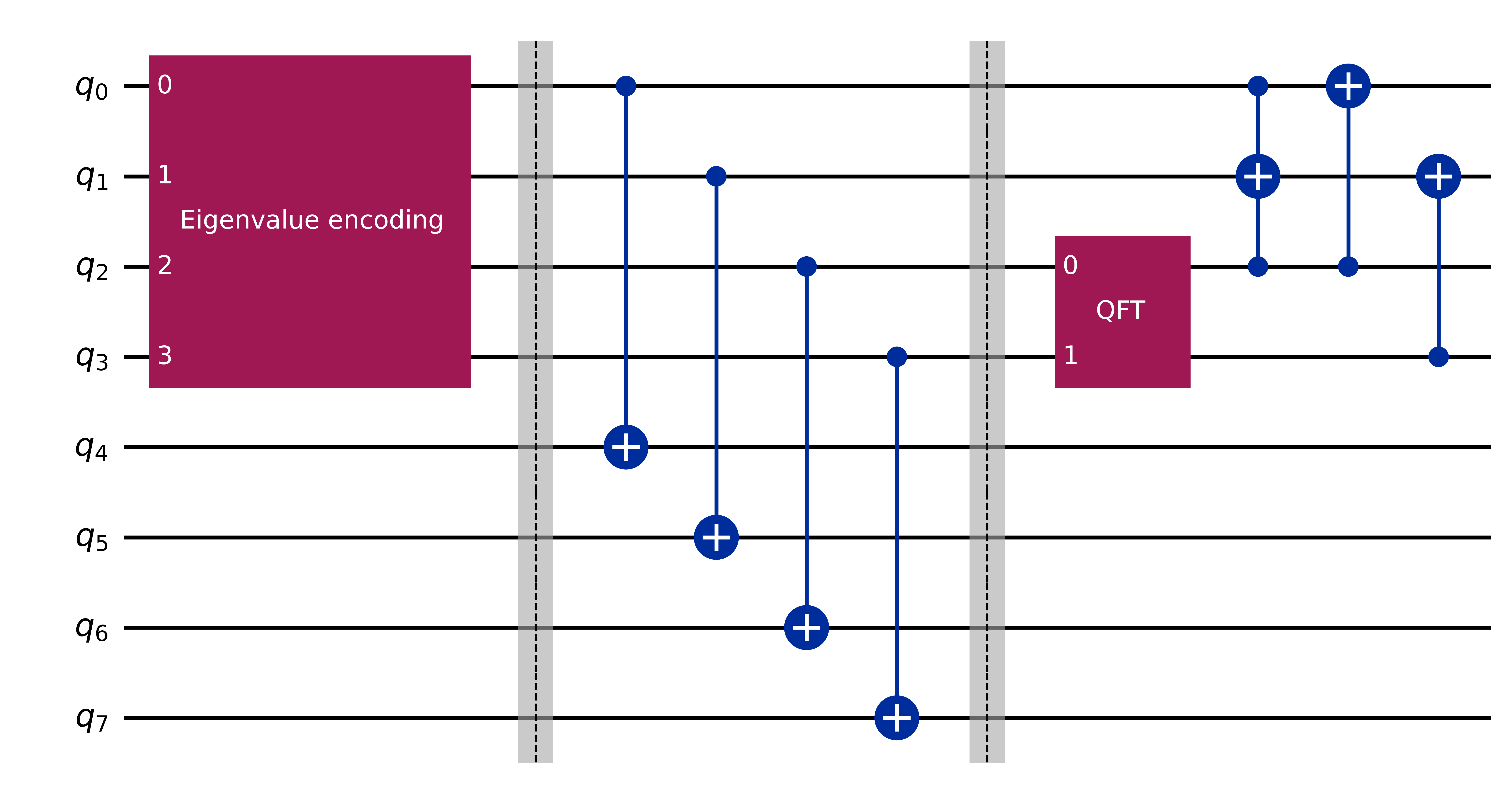}
    \caption{
    Quantum circuit for preparing two-ququart Bell diagonal states. The second block ``prepares'' the eigenvalues of a mixed two-qubit state. The third block injects entropy through entanglement with the auxiliary qubits. The fourth block changes from the computational to the basis of eigenvectors of the density operator.
    }
    \label{qc_bds_ququart}
\end{figure}
\end{center}
\end{widetext}

\subsection{Preparing arbitrary random states}
\label{subsec: random states}

In this subsection, we aim to assess the effectiveness of the mixed-state preparation algorithm in the generation of random quantum states. To this end, we adopt the fidelity between the prepared state $\sigma$ and the random target state $\rho$ as a performance metric, defined as
\begin{equation}
F(\rho,\sigma) = \left(\Tr\sqrt{\sqrt{\rho}\,\sigma\,\sqrt{\rho}}\right)^2.
\end{equation}

\begin{figure}[t]
\centering
\includegraphics[width=0.55\textwidth]{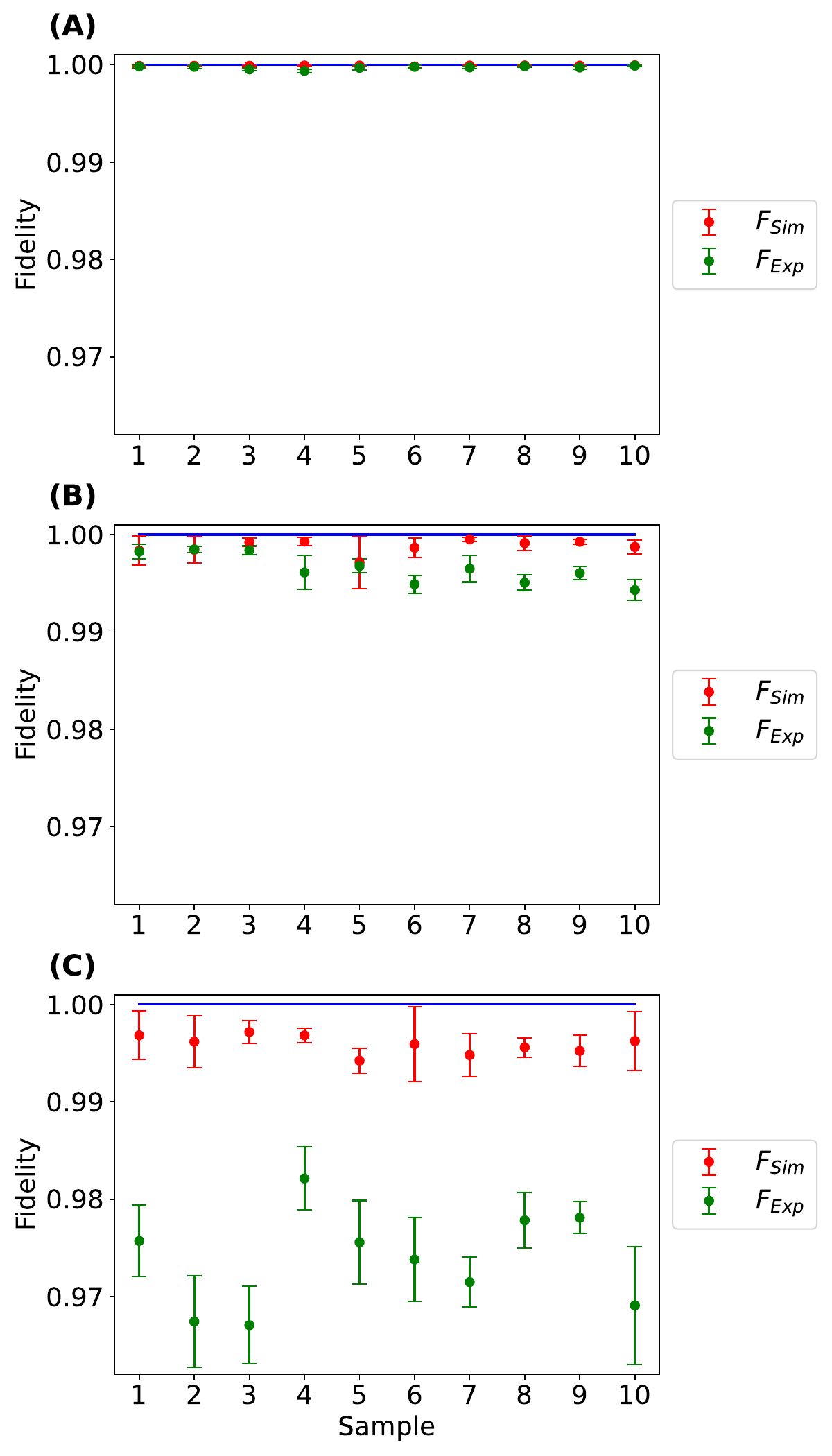}
\caption{Analysis of simulated ($F_{Sim}$) and experimental ($F_{Exp}$) fidelity for random mixed states. For each generated state, both simulated and experimental, fidelity was evaluated five times; thus, each point in the graphs represents the average fidelity and its corresponding standard deviation. In the plots are shown the fidelity for random mixed states with dimension $d=2$ in (A), dimension 
$d=4$ in (B), and dimension $d=8$ in (C). 
}
\label{fidelity_random}
\end{figure}

The generation of the random states $\rho$ is carried out using the Ginibre method \cite{bjp_rdm}. This method involves the construction of a random matrix $G \in \mathbb{C}^{d \times d}$, from which the state $\rho$ is obtained according to the following expression:
\begin{equation}
\rho = \frac{G G^{\dagger}}{\Tr[G G^{\dagger}]},
\end{equation}
where the real and imaginary parts of the matrix elements of $G$ are generated randomly with a uniform distribution in $[-1,1]$. The results obtained for random states are shown in Fig.~\ref{fidelity_random}. The obtained results, considering the numerical and statistical fluctuations and the noise in the quantum devices, attest the correct functioning of our quantum algorithm.

\section{Complexity analysis}

In this section, we provide an analysis of the resources required by our mixed state preparation algorithm. We evaluate the number of two-qubit gates (CNOTs) and the circuit depth as a function of the number \(n\) of qubits of the target state \(\rho\).

Let \(\rho\) be an arbitrary mixed state of \(n\) qubits. Our modular protocol implements a \(2n\)-qubit purification using three main blocks (see Sec.~\ref{sec:simul}), which will be analyzed separately:

\begin{itemize}
\item Eigenvalue encoding: An \(n\)-qubit pure state with real coefficients can be prepared by means of a recursive circuit employing only \(R_y\) rotations and CNOT gates. The complexity of such a construction is well known and requires \(O(2^n)\) one- and two-qubit gates 
\cite{Shende2006}.
\item Entropy injection: \(n\) CNOT gates are applied between the first \(n\) qubits (controls) and the last \(n\) qubits (targets), resulting in a total of \(n\) two-qubit gates.
\item Eigenvector preparation: This operation is an arbitrary \(n\)-qubit unitary that transforms the computational basis into the eigenvector basis of \(\rho\), requiring  \(O(2^{n/2})\) one- and two-qubit gates and \(2^{O(n)}\) ancilla \cite{Rosenthal2026}, and therefore constitutes the dominant contribution to the overall circuit complexity.
\end{itemize}

\section{Final remarks}
\label{sec:conc}

In conclusion, we have presented a novel modular quantum algorithm tailored for generating arbitrary mixed quantum states within discrete-variable quantum systems. Through experimental demonstrations, we have validated the efficacy of our algorithm in state preparation using quantum processors. Our tests encompassed the preparation of mixed states for both two-qubit X- and non X-states, along with general random density matrices spanning one, two, and three qubits. 

We investigated key properties such as quantum entanglement and quantum coherence for X- and non X-states. Additionally, the functionality of our algorithm for random states was evaluated using the quantum fidelity function, which compares the randomly generated state with the state prepared via our quantum circuit. These findings collectively underscore the 
effectiveness of our approach in preparing mixed quantum states, offering promising prospects for applications across various quantum information processing tasks. 

It is worthwhile pointing out that the computational complexity for preparing general pure quantum states, or to implement general unitary transformations, is well known to be exponential in the number of qubits of the system. The computational complexity for preparing general mixed quantum states is even higher. The same holds for our algorithm in the general case. However, the modularity of our approach (eigenvalue encoding, entropy injection, and eigenvector preparation), besides generalizing previous results from the literature, 
can stimulate the search for efficient algorithms for preparing particular classes of quantum states and allows each layer to be optimized independently in future work.
So, besides being a useful tool for studying quantum resources of low dimension quantum systems, some possibly fruitful directions for research with our algorithm are the preparation of high and low temperature Gibbs or similar states and the implementation of the eigenvalues and/or eigenvectors related unitaries using the variational quantum algorithm.

\begin{acknowledgments}
This work was supported by the Coordination for the Improvement of Higher Education Personnel (CAPES) under Grants No. 88887.829212/2023-00 and No. 88887.827989/2023-00, by the National Council for Scientific and Technological Development (CNPq) under Grants No. 300083/2025-4, No. 409673/2022-6, and No. 421792/2022-1, by the Research Support Foundation of the State of Rio Grande do Sul (FAPERGS) under Grant No. 25/2551-0002608-3, and by the National Institute for the Science and Technology of Applied Quantum Computation (INCT-CQA) under Grant No. 408884/2024-0.
\end{acknowledgments}

\vspace{0.3cm}
\textbf{Data availability.}
The data that support the findings of this study are available at \href{https://github.com/lucasfriedrich97/Mixed-state-preparation}{https://github.com/lucasfriedrich97/Mixed-state-preparation}.


\appendix

\section{Additional results for X and non-X two-qubit states}
\label{sec:appB}

\begin{figure*}
    \centering
    \includegraphics[width=0.9\linewidth]{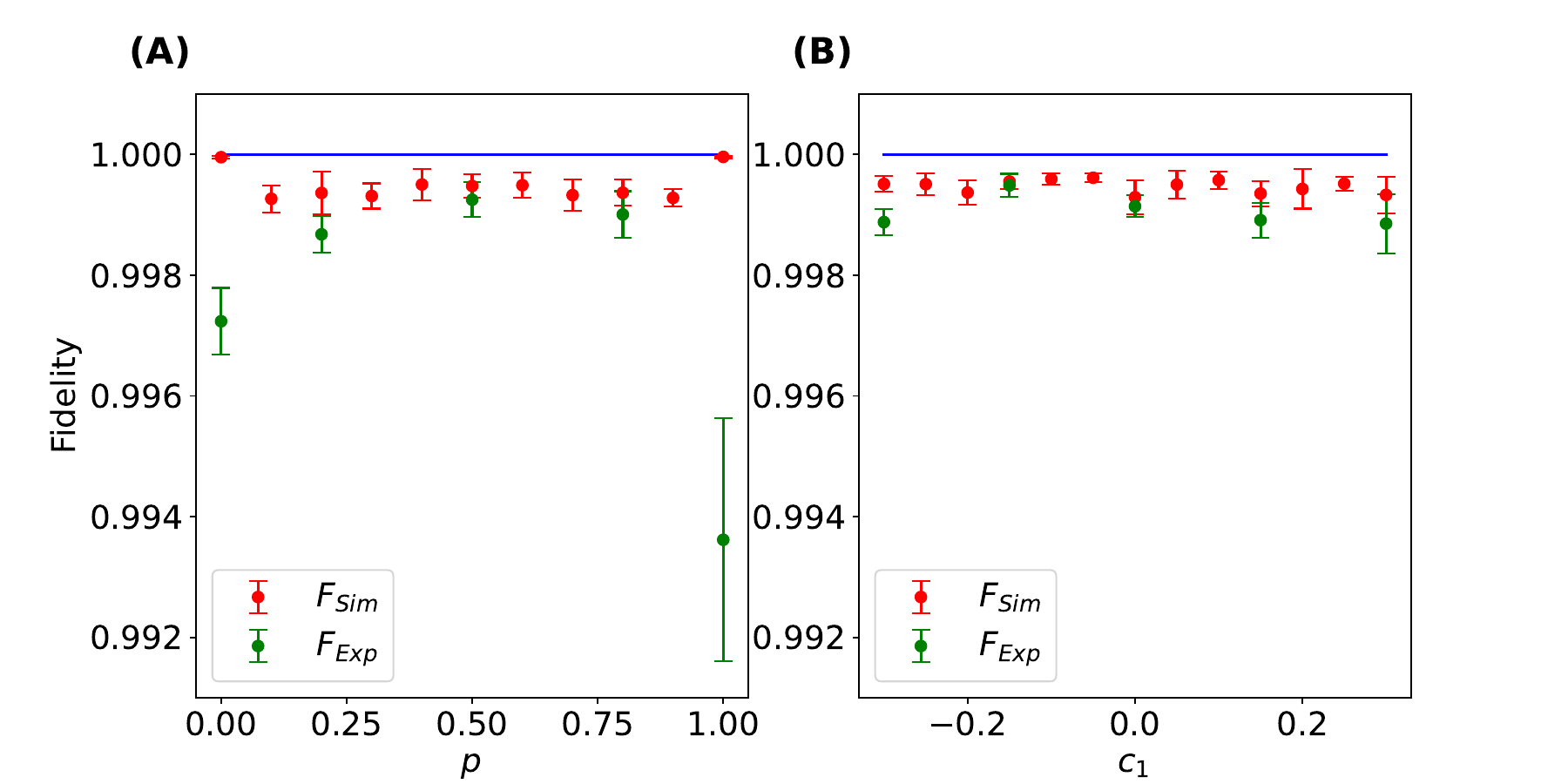}
    \caption{Fidelity of the X- and non X-states, as defined in Eqs. \eqref{eq:x_state} and \eqref{eq:non_x_state}, respectively. (A) Fidelity of the X-state, showing simulated ($F_{Sim}$) and experimental ($F_{Exp}$) results. (B) Fidelity of the non X-state, also showing simulated ($F_{Sim}$) and experimental ($F_{Exp}$) results. 
    }
    \label{fig:fidelity_X_state_non_X_state}
\end{figure*}

\begin{figure*}
    \centering
    \includegraphics[width=0.9\linewidth]{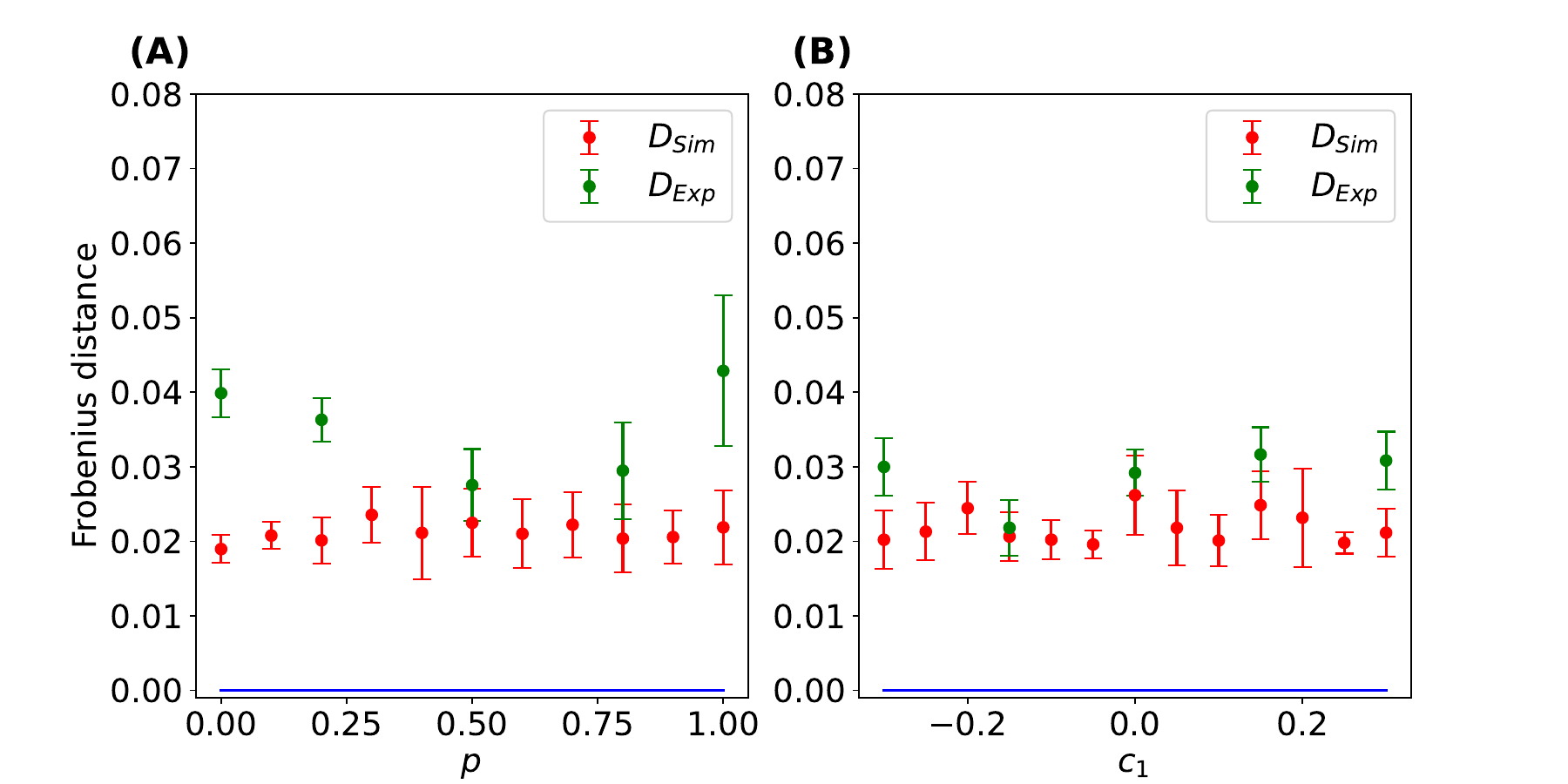}
    \caption{Frobenius distance of the X- and non X-states, as defined in Eqs. \eqref{eq:x_state} and \eqref{eq:non_x_state}, respectively. (A) Frobenius distance  of the X-state, showing simulated ($D_{Sim}$) and experimental ($D_{Exp}$) results. (B) Frobenius distance  of the non X-state, also showing simulated ($D_{Sim}$) and experimental ($D_{Exp}$) results.
    }
    \label{fig:frobenius_distance_X_state_non_X_state}
\end{figure*}

In this appendix, we provide additional results for the states prepared in this article. In Figs. \ref{fig:fidelity_X_state_non_X_state} and \ref{fig:frobenius_distance_X_state_non_X_state} are shown, respectively, the fidelity and Frobenius distance for the prepared X- and non X-states, discussed in Sec. \ref{subsec:xstates}.

\section{Quantum chip calibration data}
\label{sec:appC}

\begin{figure}
    \centering
    \includegraphics[width=0.3\linewidth]{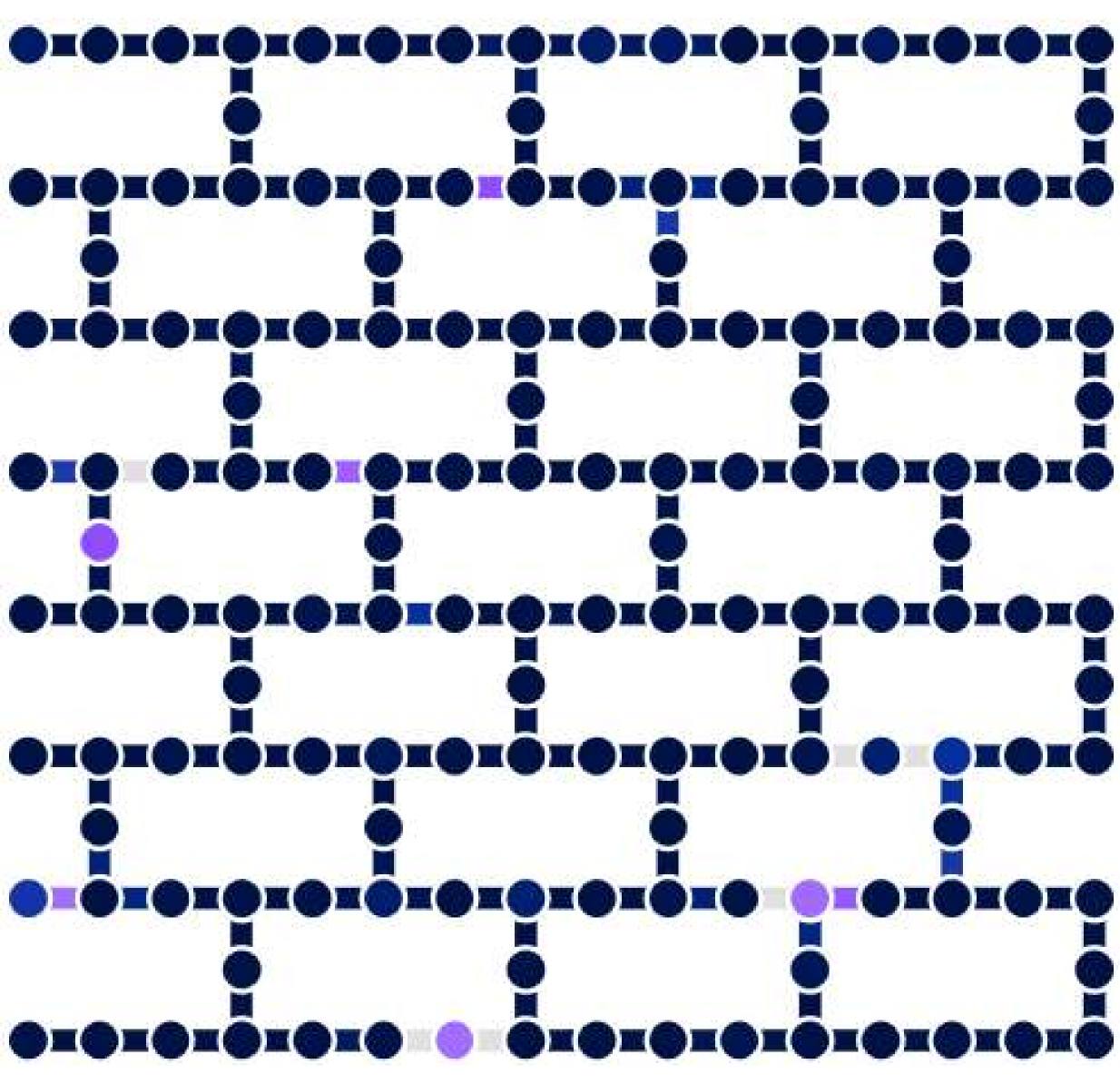}
    \caption{This figure shows the connections architecture of IBMQ's quantum chip: \textit{ibm\_kingston}.}
    \label{fig:chip_ibm_kingston}
\end{figure}

In our experiments, we used the IBM Quantum chip \textit{ibm\_kingston}. The chip architecture is shown in Fig.~\ref{fig:chip_ibm_kingston} and its calibration data are shown in Table \ref{cal_data}.

\begin{table}[ht]
\centering
\begin{tabular}{c|c|c|c|c}
\toprule
CZ error & SX error & Readout error & T1($\mu s$) & T2 ($\mu s$) \\
\midrule
1.986E-3 & 2.335E-4 & 7.446E-3 & 278.49  & 157.32 \\
\bottomrule
\end{tabular}
\caption{Calibration data for IBMQ's \textit{ibm\_kingston} quantum chip. In all cases, the mean value is shown.}
\label{cal_data}
\end{table}

\begin{thebibliography}{10}
\bibliographystyle{apsrev4-1}



\bibitem[]{Feynman1982} R. P. Feynman, Simulating physics with computers, Int. J. Theor. Phys. \textbf{21}, 467 (1982).


\bibitem[]{Benioff1982} P. Benioff, Quantum Mechanical Models of Turing Machines That Dissipate No Energy, Phys. Rev. Lett. \textbf{48}, 1581 (1982).


\bibitem[]{Deutsch1985} D. Deutsch, Quantum theory, the Church-Turing principle and the universal quantum computer, Proc. R. Soc. Lond. A \textbf{400}, 1818 (1985).


\bibitem[]{Steane1988} A. Steane, Quantum computing, Rep. Prog. Phys. \textbf{61}, 117 (1998).


\bibitem[]{Ladd2010} T. D. Ladd, F. Jelezko, R. Laflamme, Y. Nakamura, C. Monroe, and J. L. O'Brien, Quantum computers, Nature \textbf{464}, 45 (2010).


\bibitem[]{OBrien2007} J. L. O'Brien, Optical Quantum Computing, Science \textbf{318}, 1567 (2007).


\bibitem[]{Mohseni2024} M. Mohseni \textit{et al.}, How to Build a Quantum Supercomputer: Scaling from Hundreds to Millions of Qubits, arXiv:2411.10406 (2024).


\bibitem[]{Girvin2023}S. M. Girvin, Introduction to quantum error correction and fault tolerance, SciPost Phys. Lect. Notes \textbf{70}, (2023).


\bibitem[]{Acharya2025} R. Acharya \textit{et al.}, Quantum error correction below the surface code threshold, Nature \textbf{638}, 920 (2025).


\bibitem[]{Putterman2025} H. Putterman \textit{et al.}, Hardware-efficient quantum error correction via concatenated bosonic qubits, Nature \textbf{638}, 927 (2025).


\bibitem[]{Reichardt2025} B. W. Reichardt \textit{et al.}, Fault-tolerant quantum computation with a neutral atom processor, arXiv:2411.11822 (2025).


\bibitem[]{Biswas2017} R. Biswas \textit{et al.}, A NASA perspective on quantum computing: Opportunities and challenges, Parallel Computing \textbf{64}, 81 (2017).


\bibitem[]{Acampora2025} G. Acampora \textit{et al.}, Quantum computing and artificial intelligence: status and perspectives, arXiv:2505.23860 (2025).


\bibitem[]{Cao2019} Y. Cao \textit{et al.}, Quantum Chemistry in the Age of Quantum Computing, Chem. Rev. 
\textbf{119}, 10856 (2019).


\bibitem[]{Ollitrault2021} P. J. Ollitrault, A. Miessen, and I. Tavernelli, Molecular Quantum Dynamics: A Quantum Computing Perspective, Acc. Chem. Res. \textbf{54}, 4229 (2021).


\bibitem[]{Liu2022} H. Liu, G. H. Low, D. S. Steiger, T. H\"aner, M. Reiher, and M. Troyer, Prospects of quantum computing for molecular sciences, Mater. Theory \textbf{6}, 11 (2022).


\bibitem[]{Baiardi2023} A. Baiardi, M. Christandl, and M. Reiher, Quantum Computing for Molecular Biology, ChemBioChem \textbf{24}, e202300120 (2023).


\bibitem[]{Meglio2024} A. Di Meglio \textit{et al.}, Quantum Computing for High-Energy Physics: State of the Art and Challenges, PRX Quantum \textbf{5}, 037001 (2024).


\bibitem[]{Tennie2025} F. Tennie, S. Laizet, S. Lloyd, and L. Magri, Quantum computing for nonlinear differential equations and turbulence, Nat. Rev. Phys. \textbf{7}, 220 (2025).


\bibitem[]{Childs2010} A. M. Childs and W. van Dam, Quantum algorithms for algebraic problems, Rev. Mod. Phys. \textbf{82}, 1 (2010).


\bibitem[]{Bharti2022} K. Bharti \textit{et al.}, Noisy intermediate-scale quantum algorithms, Rev. Mod. Phys. \textbf{94}, 015004 (2022).


\bibitem[]{Santoro2006} G. E. Santoro and E. Tosatti, Optimization using quantum mechanics: quantum annealing through adiabatic evolution, J. Phys. A \textbf{39}, R393 (2006).


\bibitem[]{Dalzell2025} A. M. Dalzell \textit{et al.}, Quantum algorithms: A survey of applications and end-to-end complexities, Cambridge (2025).


\bibitem[]{Bauer2020} B. Bauer, S. Bravyi, M. Motta, and G. K.-L. Chan, Quantum Algorithms for Quantum Chemistry and Quantum Materials Science, Chem. Rev. \textbf{120}, 12685 (2020).


\bibitem{Montanaro2016} A. Montanaro, Quantum algorithms: An overview, npj Quantum Inf. \textbf{2}, 15023 (2016).

\bibitem{qiskit} Qiskit contributors,
Qiskit: An Open-source Framework for Quantum Computing, 2023, 10.5281/zenodo.2573505.


\bibitem{ibmq} IBM Quantum Plataform, \href{https://quantum-computing.ibm.com/}{https://quantum-computing.ibm.com/}.


\bibitem[]{Plesch2011} M. Plesch and \v{C}. Brukner, Quantum-state preparation with universal gate decompositions, Phys. Rev. A \textbf{83}, 032302 (2011).


\bibitem[]{Cruz2019} D. Cruz \textit{et al.}, Efficient Quantum Algorithms for GHZ and W States, and Implementation on the IBM Quantum Computer, Adv. Quan. Tech. \textbf{2}, 1900015 (2019).


\bibitem[]{Araujo2021} I. F. Araujo, D. K. Park, F. Petruccione, and A. J. da Silva, A divide-and-conquer algorithm for quantum state preparation, Sci. Rep. \textbf{11}, 1 (2021).


\bibitem[]{Zhang2022} X.-M. Zhang, T. Li, and X. Yuan, Quantum State Preparation with Optimal Circuit Depth: Implementations and Applications, Phys. Rev. Lett. \textbf{129}, 230504 (2022).


\bibitem[]{Yeo2025} H. Yeo, H. E. Kim, I. Sohn, and K. Jeong, Reducing circuit depth in quantum state preparation for quantum simulation using measurements and feedforward, Phys. Rev. Appl. \textbf{23}, 054066 (2025).


\bibitem[]{Yin2025} C. Yin, Fast and Accurate Greenberger-Horne-Zeilinger Encoding Using All-to-All Interactions, Phys. Rev. Lett. \textbf{134}, 130604 (2025).


\bibitem[]{Yuan2023} P. Yuan and S. Zhang, Optimal (controlled) quantum state preparation and improved unitary synthesis by quantum circuits with any number of ancillary qubits, Quantum \textbf{7},  956 (2023).


\bibitem[]{Perdomo2025} O. Perdomo, N. Castaneda, and R. Vogeler, Preparation of three-qubit states, Int. J. Quantum Inform. \textbf{23}, 2450046 (2025).


\bibitem{kitaev1995} A. Y. Kitaev, Quantum measurements and the Abelian Stabilizer Problem, 	arXiv:quant-ph/9511026 (1995).


\bibitem{Shende2006} V. V. Shende, S. S. Bullock, and I. L. Markov, Synthesis of Quantum Logic Circuits, IEEE Trans. on Computer-Aided Design \textbf{25}, 1000 (2006).


\bibitem{Arrazola2019} J. M. Arrazola, T. R. Bromley, J. Izaac, C. R. Myers, K. Br\'adler, and N. Killoran, Machine learning method for state preparation and gate synthesis on photonic quantum computers, Quantum Sci. Technol. \textbf{4}, 024004 (2019).


\bibitem{He2021} R.-H. He, H.-D. Liu, S.-B. Wang, J. Wu, S.-S. Nie, and Z.-M. Wang, Universal quantum state preparation via revised greedy algorithm, Quantum Sci. Technol. \textbf{6}, 045021 (2021).


\bibitem{Zhang2021} X.-M. Zhang, M.-H. Yung, and X. Yuan, Low-depth quantum state preparation, Phys. Rev. Res. \textbf{3}, 043200 (2021).


\bibitem{Veras2022} T. M. L. de Veras, L. D. da Silva, and A. J. da Silva, Double sparse quantum state preparation, Quantum Inf. Process. \textbf{21}, 204 (2022).


\bibitem{Xin2017} T. Xin, S.-J. Wei, J. S. Pedernales, E. Solano, and G.-L. Long, Quantum simulation of quantum channels in nuclear magnetic resonance, Phys. Rev. A \textbf{96}, 062303 (2017).


\bibitem{Wei2018} S.-J. Wei, T. Xin, and G.-L. Long, Efficient universal quantum channel simulation in IBM's cloud quantum computer, Sci. China Phys. Mech. Astron. \textbf{61}, 70311 (2018).


\bibitem{Zanetti2023} 
M. S. Zanetti, D. F. Pinto, M. L. W. Basso, and J. Maziero, Simulating noisy quantum channels via quantum state preparation algorithms, J. Phys. B \textbf{56}, 115501 (2023).


\bibitem{Clemente2024} G. Clemente, Mixed State Variational Quantum Eigensolver for the Estimation of Expectation Values at Finite Temperature,
arXiv:quant-ph/2401.17194 (2024).


\bibitem{Yordanov2019} Y. S. Yordanov and C. H. W. Barnes, Implementation
of a general single-qubit positive operator-valued measure on a circuit-based quantum computer, Phys. Rev. A \textbf{100}, 062317 (2019).


\bibitem{Pinto2023}
D. F. Pinto, M. S. Zanetti, M. L. Basso and J. Maziero, Simulation of positive operator-valued measures and quantum instruments via quantum state-preparation algorithms. Phys. Rev. A \textbf{107}, 022411 (2023).


\bibitem{Caves2010}
M. D. Lang and  C. M. Caves, Quantum Discord and the Geometry of Bell-Diagonal States, Phys. Rev. Lett. \textbf{105}, 150501 (2010).


\bibitem{Pozzobom2019} M. B. Pozzobom and J. Maziero, Preparing tunable Bell-diagonal states on a quantum computer, Quantum Inf. Process. \textbf{18}, 142 (2019).


\bibitem{Garding2021} E. R. G\r{a}rding et al., Bell Diagonal and Werner State Generation: Entanglement, Non-Locality, Steering and Discord on the IBM Quantum Computer, Entropy \textbf{23}, 797 (2021).


\bibitem{Shahbeigi2022} F. Shahbeigi and M. Karimi and V. Karimipour, Simulating of X-states and the two-qubit XYZ Heisenberg system on IBM quantum computer, Physica Scripta \textbf{97}, 025101 (2022).


\bibitem{Falco2026} 
A. Falco, D. Falco-Pomares, and H. G. Matthies, A Rigorous and Self--Contained Proof of the Grover-Rudolph State Preparation Algorithm,
arXiv:2601.17930 (2026).


\bibitem{Alhajjar2023} 
E. Alhajjar, J. Geneson, A. Prakash, and N. Robles, Efficient quantum loading of probability distributions through Feynman propagators, arXiv:2311.13702 (2023).


\bibitem{Grover2002} 
L. Grover and T. Rudolph, Creating superpositions that correspond to efficiently integrable probability distributions,
arXiv:quant-ph/0208112 (2002).


\bibitem[]{Lomwel2026} 
A. van Lomwel, P. M. Schindler, M. Orozco-Ruiz, M. Bukov, N. H. Le, and F. Mintert, Fast thermal state preparation beyond native interactions,
arXiv:2601.04810 (2026).


\bibitem[]{Rouze2026} 
C. Rouz\'e, D. S. Fran\c{c}a, and \'A. M. Alhambra, Optimal quantum algorithm for Gibbs state preparation, Phys. Rev. Lett. \textbf{136}, 060601 (2026).

\bibitem{Nielsen2000} 
M.~A.~Nielsen and I.~L.~Chuang, \textit{Quantum Computation and Quantum Information}, Cambridge University Press (2010).

\bibitem[]{Castro2016} C. S. Castro, O. S. Duarte, D. P. Pires, D. O. Soares-Pinto, and M. S. Reis, Thermal entanglement and teleportation in a dipolar interacting system, Phys. Lett. A \textbf{380}, 1571 (2016).


\bibitem{Wootters2001} W. K. Wootters,  Entanglement of formation and concurrence. Quantum Inf. Comput. \textbf{1}, 27 (2001).

\bibitem[]{Zhang2005} G.-F. Zhang and S.-S. Li, Thermal entanglement in a two-qubit Heisenberg $XXZ$ spin chain under an inhomogeneous magnetic field, Phys. Rev. A \textbf{72}, 034302 (2005).

    
\bibitem{bjp_rdm} J. Maziero, Random Sampling of Quantum States: a Survey of Methods, Braz. J. Phys. \textbf{45}, 575 (2015).



\bibitem{Rosenthal2026} G. Rosenthal, Query and Depth Upper Bounds for Quantum Unitaries via Grover Search, Quantum 10, 2144 (2026).



\end{thebibliography}
\end{document}